\documentclass[a4paper]{article}

\usepackage{INTERSPEECH2022}
\usepackage{microtype}
\usepackage{graphicx}
\usepackage{booktabs} 
\usepackage{amsmath}
\usepackage{soul}
\usepackage{booktabs}
\usepackage{algorithm}
\usepackage{algorithmic}
\usepackage{subcaption}
\usepackage{amsfonts}
\usepackage{multirow}
\usepackage{hyperref}
\usepackage{soul}

\title{Regotron: Regularizing the Tacotron2 architecture via monotonic alignment loss}
\name{Efthymios Georgiou$^{1,2}$, Kosmas Kritsis$^1$, Georgios Paraskevopoulos$^{1,2}$, Athanasios Katsamanis$^1$, Vassilis Katsouros$^1$, Alexandros Potamianos$^2$}
\address{
  $^1$Institute for Language and Speech Processing, Athena Research Center, Athens, Greece\\
  $^2$School of ECE, National Technical University of Athens, Athens, Greece}
\email{\{\footnotesize{e.georgiou,kosmas.kritsis,g.paraskevopoulos,nkatsam,vsk\}}@athenarc.gr, potam@central.ntua.gr}

\begin{document}

\maketitle
\begin{abstract}
Recent deep learning Text-to-Speech (TTS) systems have achieved impressive performance by generating speech close to human parity. However, they suffer from training stability issues as well as incorrect alignment of the intermediate acoustic representation with the input text sequence. In this work, we introduce Regotron, a regularized version of Tacotron2 which aims to alleviate the training issues and at the same time produce monotonic alignments. Our method augments the vanilla Tacotron2 objective function with an additional term, which penalizes non-monotonic alignments in the location-sensitive attention mechanism. By properly adjusting this regularization term we show that the loss curves become smoother, and at the same time Regotron consistently produces monotonic alignments in unseen examples even at an early stage (13\% of the total number of epochs) of its training process, whereas the fully converged Tacotron2 fails to do so. 
Moreover, our proposed regularization method has no additional computational overhead, while reducing common TTS mistakes and achieving slighlty improved speech naturalness according to subjective mean opinion scores (MOS) collected from 50 evaluators.
\end{abstract}
\noindent\textbf{Index Terms}: speech synthesis, regularization, alignment, Tacotron2, Regotron

\section{Introduction}
Text-to-Speech (TTS) is the task in which the machine is challenged to generate speech from text. Successfully building such models, is of vital importance for any realistic human machine interaction system.
TTS has therefore a plethora of applications, which may vary from typical phone assistants, to apps suited for people with speaking disorders.

Neural TTS systems ~\cite{wavenet, tacotron2} have achieved impressive performance, by generating natural speech close to human parity. Most systems usually incorporate two modules. The first is utilized to map the input text sequence to some acoustic features ~\cite{tacotron, deepvoice3}, e.g., mel spectrograms, while the second transforms the predicted features to speech waveforms ~\cite{parallel-wavenet, wavernn, waveglow}. The latter module is also called a vocoder. Some approaches in the literature also aim to tackle the problem in an end-to-end manner ~\cite{deepmind_tts_2021}.

The predominant architecture for the text to mel mapping is a sequence-to-sequence model ~\cite{sutskever2014sequence} based on the encoder-decoder setting ~\cite{tacotron, deepvoice3}. The alignment module which is responsible for aligning the input text with the intermediate mel representations, is usually the cornerstone ~\cite{tacotron2, clarinet, transformer-tts, ren2019fastspeech} of these approaches.

Finding proper alignments assists the network's convergence during training and additionally encourages the model to generate more natural spectrograms. Inaccurate alignments may therefore result in non-convergence issues and this is the reason many non-autoregressive approaches require a pretrained autoregressive teacher network ~\cite{ren2019fastspeech, transformer-tts, fastspeech-2, parallel_taco2}. Additionally, common TTS mistakes such as repetitions and word skips are attributed to the alignment mechanism ~\cite{ren2019fastspeech}. Thus, for any TTS mel-generating architecture it is crucial to learn how to properly align the input text with the mel-spectrogram.  

A fruitful line of research is therefore to improve the training stability and alignments of such text-to-mel models. Various approaches have been proposed in the literature. For example, Flowtron ~\cite{flowtron} progressively trains and stacks flows to stabilize training, while ~\cite{okamoto} uses an acoustic model as force aligner for the alignment module. In ~\cite{chung21_interspeech} authors utilize a reinforcement learning framework where the agent is a duration predictor network. FlowTTS ~\cite{flow-tts} also uses a length predictor to stabilize the training of consecutive normalizing flows.

More similar to our work, ~\cite{dou21_interspeech} assumes a prior Laplacian distribution on the weights and aims to achieve diagonal alignments. Battenberg et. al. ~\cite{battenberg2020location} also identify the alignment issues in the Tacotron2 architecture, and utilize two families of attention modules to address this issue, such as GMM-based mechanisms.
In ~\cite{lin21g_interspeech} authors use alignments from independent trainable auxiliary attention mechanisms to better guide the main alignment module. The most popular ideas however, appear to be iterative-based and specifically works that use Viterbi-like algorithms in order to pick the optimal (monotonic) across multiple alignments ~\cite{glow-tts, aligntts, one-to-rule-tem}. However, all of the aforementioned approaches, either use restrictive assumptions, or introduce significant computational overhead.

In this work, we take a different approach and directly address both training stability and alignment issues, by adding a regularization term in the total objective function of the Tacotron2 architecture. 
This way, no additional computational overhead is introduced.
Formally, we augment the loss function with a term that acts as a regularizer and imposes the monotonicity requirement, by penalizing non-monotonic neighboring alignment weights ~\cite{rios-etal-2021-biasing}. The regularized architecture we propose is called \textit{Regotron}. Our experiments verify that Regotron has smoother training behavior compared to Tacotron2 and 
consistently produces monotonic
alignments at an early stage (13\% of total epochs) of its training process, in unseen examples, where the fully converged
Tacotron2 fails to do so. Moreover, the additional term has a beneficial effect on the overall loss, since Regotron achieves lower generalization error than Tacotron2. 
Additionally, we experimentally verify that common TTS mistakes in difficult examples are reduced when our regularization is used. Finally, a MOS evaluation shows that Regotron achieves slightly improved speech naturalness compared to Tacotron2.

Our contributions can be summarized as: 
1) We introduce an efficient regularization term which acts directly on the alignment weights, enforcing monotonicty, with no additional architectural modifications or significant computation cost 2) We show that this leads to enhanced training stability and generalization due to better regularization, 3) We finally demonstrate that increased robustness to common mistakes is achieved by just adding this regularization term. Our code is available as open-source \footnote{link to be provided upon acceptance}.

\section{Proposed Method}

\subsection{Tacotron-2}
Tacotron2 is an autoregressive encoder-decoder architecture equipped with a location-sensitive attention mechanism. The encoder takes as input text characters and outputs hidden representations which are fed to the decoder, which in turn generates the mel spectrogram. An important part of the architecture is the location-sensitive attention mechanism, which helps the decoder attend at different parts of the encoder's hidden representations. The resulting matrix of alignment scores (as illustrated in Fig.~\ref{fig:alignments}), describes how well a part of the input, i.e., a character (vertical axis), is aligned with the corresponding generated mel frames (horizontal axis). We formally describe the alignment matrix as $A \in \mathbb{R}^{N\times M}$, where $N$ is the number of input characters and $M$ the number of the generated mel frames. In other words $a_{ij}$ is the probability weight that informs us how well the $i$-th input character is aligned with the $j$-th mel frame. Note here that TTS requires monotonic alignment, which means that if the $i$-th input character is mapped to the $j$-th mel frame then the $(i+1)$-th input character should be mapped at frame $k > j$.

\subsection{Monotonic Alignment Loss}
In order to effectively encode the monotonic alignment requirement we aim for an objective which captures monotonicity ~\cite{rios-etal-2021-biasing}. Given the alignment matrix $A \in \mathbb{R}^{N\times M}$, we follow ~\cite{rios-etal-2021-biasing} and define the ``mean attended position'' (or centroid position) as:
\begin{equation}\label{eq:centroid}
    \langle a_j \rangle = \sum_{i=1}^N a_{ij} \cdot i
\end{equation}
which calculates the centroid of the alignment weight for a particular mel-frame $j$.
We then use these centroids to encode the monotonicity requirement by setting:
    $\langle a_{j+1} \rangle \geq \langle a_{j} \rangle$, where $\quad j \in \{1, \cdots M\}$ 
,which describes that the $j+1$-th mean attended position should be larger (or equal) than the $j$-th. ~\cite{rios-etal-2021-biasing} modifies this requirement 
by introducing a hyperparameter $\delta$ which is dynamically reweighted by the the fraction of input and output sequence lengths. Putting it all together the alignment objective can be expressed as in ~\cite{rios-etal-2021-biasing}:
\begin{equation}\label{eq:align_loss}
    L_{A} = \sum_{j=1}^{M-1} \max\Big\{\frac{\langle a_{j} \rangle - \langle a_{j+1} \rangle + \delta\frac{N}{M}}{N}, 0 \Big\}
\end{equation}
where $\delta$ controls how much to amplify the monotonicity assumption, $N/M$ dynamically adjusts $\delta$ according to input and output sequence lengths and finally the denominator $N$ acts as an input-based ``normalization" factor. If the fist term in the max operand becomes negative, i.e., the monotonicity requirement is satisfied, then it is neglected from the total sum in Eq.~\eqref{eq:align_loss}, thus penalizing only the terms that violate the monotonicity demand.

\subsection{Regotron}
The proposed architecture, called Regotron, is a regularized version of Tacotron2. Essentially, we enforce the location-sensitive attention mechanism of Tacotron2 to find monotonic alignments, by adding an extra loss term. This addition, stabilizes the training procedure of Tacotron2 and also gets correct alignments earlier during the training procedure. Moreover, it introduces no additional computational overhead. 
Regotron objective function $L_R$ is therefore the Tacotron2 objective $L_T$ augmented by the regularization term $L_A$ of Eq.~\ref{eq:align_loss} which penalizes non-monotonic alignments:
\begin{equation}
    L_{R} = L_{T} + \lambda L_{A}
\end{equation}
where $\lambda$ is a positive weighting value, used to adjust the effect of the second (alignment) term. In practice we only tune this hyperparameter and not $\delta$ (see Section 3). Note here that the second term is applied directly to the attention matrix $A$, while the first to the generated mels.

\section{Experimental Setup}\label{sec:exp_setup}
We follow the standard practice ~\cite{tacotron2, waveglow} and train our models using the LJSpeech (LJS) dataset ~\cite{ljspeech17}.
We use a sampling rate of 22050 Hz and mel-spectrograms
with 80 bins. 
We apply the
STFT with a FFT size of 1024, hop size of 256 ($\sim$12ms), and window
size of 1024 samples. Following the implementation from nvidia \footnote{https://github.com/NVIDIA/DeepLearningExamples/tree/master
/PyTorch/SpeechSynthesis/Tacotron2} we choose Adam optimizer with learning rate of $10^{-3}$ and train the model for $1500$ epochs. We also anneal the learning rate by a factor of $0.3$ every $500$ epochs. For a fair comparison, we retrain Tacotron2 and the results refer to our version. All models are trained using mixed precision for faster training and a batch size of $154$. For any additional hyperparameters, we refer to nvidia implementation. For the Regotron architecture we use the exact same hyperparameters as in Tacotron2. We additionally set $\delta=0.01$ for all experiments and we only perform a hyperparameter search for $\lambda$ in the range $10^{-6}$ to $10^{-2}$, with (multiplicative) step $10^{-1}$. For our setup we find $10^{-5}$ to perform better.

\section{Experiments}

\subsection{Loss function analysis}
\begin{figure*}[!htb]
     \centering
     \begin{subfigure}[!h]{0.48\linewidth}
         \centering
        \includegraphics[width=\linewidth]{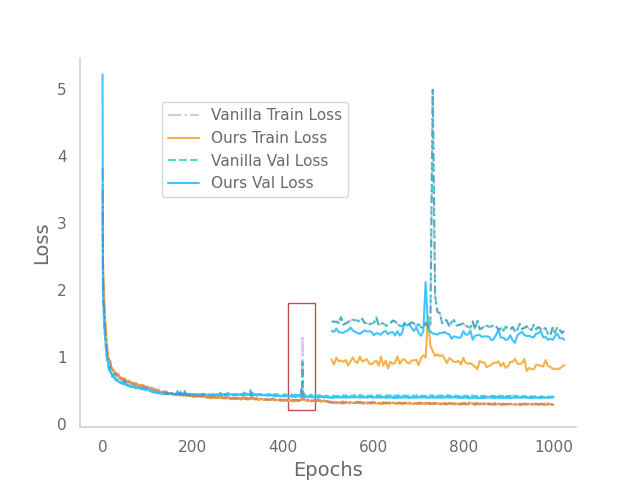}
        \caption{Loss functions up to epoch 1000}
         \label{fig:up2-1000}
     \end{subfigure}
     \hfill
     \begin{subfigure}[!h]{0.48\linewidth}
        \centering
        \includegraphics[width=\linewidth]{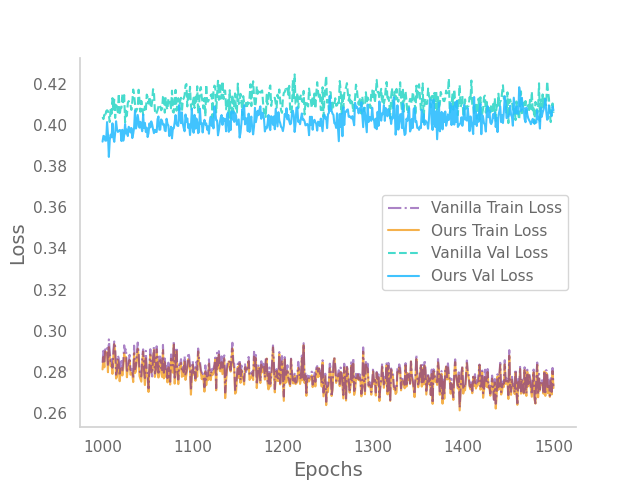}
        \caption{Loss functions from epoch 1000 to 1500}
         \label{fig:from-1000}
     \end{subfigure}
     \caption{
     Train and validation loss functions for the Tacotron2 (Vanilla) as well Regotron (Ours). The red box in the left figure is zoomed in the center right of the same image.
     }
     \label{fig:Loss functions}
\end{figure*}

In this section we analyze the loss function behavior during training.
Specifically, we illustrate the respective error $L_{T}$ for both the Tacotron2 (Vanilla), as well as Regotron (Ours) in Fig~\ref{fig:Loss functions}. 

\subsubsection{Training Stabilization}
Fig~\ref{fig:up2-1000} shows that Regotron (Ours) does not appear to have spiky behavior, while Tacotron2 (Vanilla) appears to do so. Since the only essential difference of Regotron and Tacotron2 is the regularization term, we attribute the spiky behavior of Tacotron2 to the lack of robustness in the alignment procedure, which is also in agreement with existing works ~\cite{battenberg2020location}. Specifically, we can see that Vanilla Train Loss as well as Vanilla Val Loss are the only errors which are spiky, see red box in Fig.~\ref{fig:up2-1000} and its zoomed version on the same image. Note here that the purpose of the zoomed version is to make clear which losses pose spiky behavior and which not. It is not meant to encapsulate the whole spike. It is therefore clear that our regularized version better \textit{stabilizes the training procedure} when weighted appropriately, i.e., for $\lambda \in \{10^{-3}, 10^{-4}, 10^{-5}\}$. For smaller weight values, e.g, $10^{-6}$, the alignment term becomes insignificant and the Regotron objective tends to the vanilla Tacotron2 one. For larger values, e.g. $10^{-2}$, the alignment term $L_A$ dominates over $L_T$, resulting in non-convergence. 

\begin{table}[!htb]
  \caption{Generalization and Validation Error averaged over the last 100 (1400-1500) and 500 (1000-1500) training epochs.}
  \label{tab:generalization_error}
  \centering
  \scalebox{0.9}{
  \begin{tabular}{ ccccc }
    \toprule
    \textbf{Model} (avg. epochs) & Val Error & Generalization \\
    \midrule
    Tacotron2 (1400-1500) & $0.4121$ & $0.1319$ \\
    Regotron (1400-1500) & $\mathbf{0.4017}$ & $\mathbf{0.1232}$ \\
    \midrule
    Tacotron2 (1000-1500) & $0.4115$ & $0.1324$ \\
    Regotron (1000-1500) & $\mathbf{0.4023}$ & $\mathbf{0.1247}$  \\
    \bottomrule
  \end{tabular}}
\end{table}
\subsubsection{Generalization Error}
Fig.~\ref{fig:from-1000} illustrates the training and validation losses from epoch 1000 till the final epoch 1500. It is clear that Regotron has lower validation error compared to the vanilla Tacotron2. This means that the generated spectrograms, for the validation set, are more similar to the ground truths (similarity is measured via $L_{T}$).
An alternative metric, is to measure the gap between the validation and the training loss for each model (generalization error). In order to so, we averaged over the last 100 and the last 500 training epochs for both models. The results are shown in Table~\ref{tab:generalization_error}. As it is clear, Regotron achieves both lower validation and generalization error, which implies that introducing the alignment loss has a beneficial effect on $L_{T}$ and results in a better performing model (in terms of the objective).

\subsection{Alignment Analysis}
\begin{figure*}[!htb]
     \centering
     \begin{subfigure}[!h]{0.33\linewidth}
         \centering
        \includegraphics[width=\linewidth]{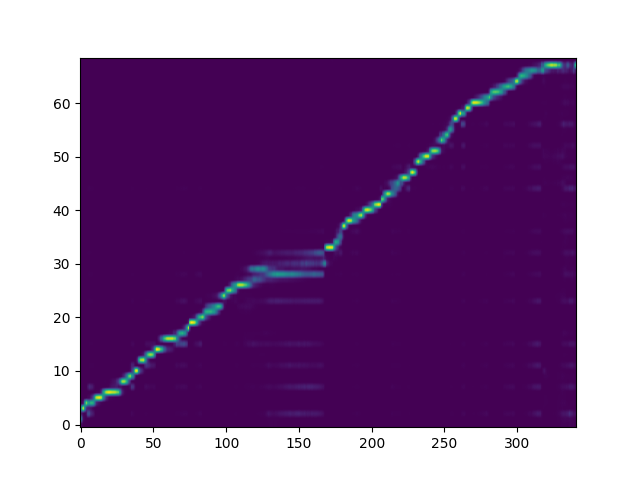}
        \caption{Tacotron2}
         \label{fig:taco-a}
     \end{subfigure}
     \hfill
     \begin{subfigure}[!h]{0.33\linewidth}
        \centering
        \includegraphics[width=\linewidth]{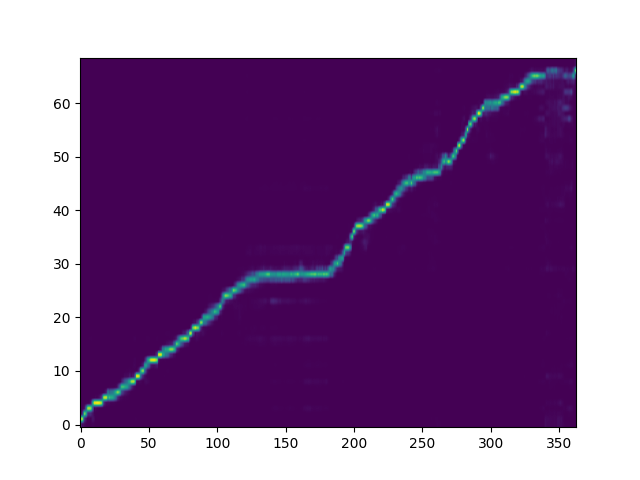}
        \caption{Regotron ($13\%$)}
         \label{fig:rego-200-a}
     \end{subfigure}
     \hfill
     \begin{subfigure}[!h]{0.33\linewidth}
        \centering
        \includegraphics[width=\linewidth]{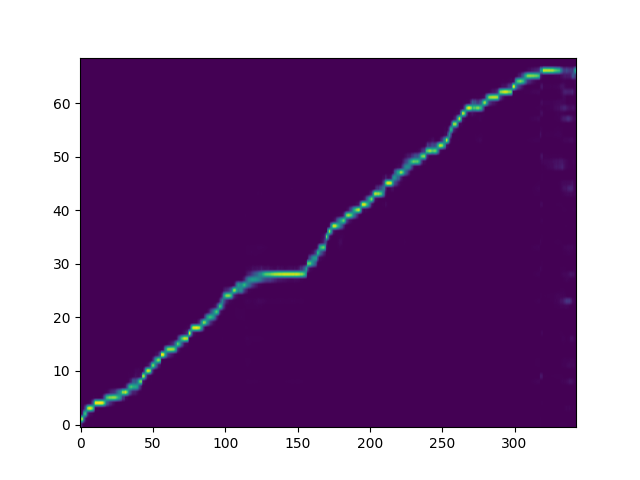}
        \caption{Regotron}
         \label{fig:rego-a}
     \end{subfigure}
     \caption{
    The plots illustrate alignments from the Tacotron2 model, an early ($13\%$ of the total training epochs) Regotron version and the fully converged Regotron. The phrase uttered is ``The jury did not believe him, and the verdict was for the defendants.''.
     }
     \label{fig:alignments}
\end{figure*}

In this section we analyze the alignment results from Regotron on phrases from the held-out test set. Since our model directly optimizes the monotonic alignment we expect to see different behavior from the vanilla Tacotron2. After inspecting our checkpoints we find that a very early version of our model 
achieves better alignment in general than the vanilla Tacotron2. This result is illustrated in Fig.~\ref{fig:alignments} where the first column denotes Tacotron2 alignments, the second shows Regotron (at $13\%$ of the total number of training epochs) alignments and the
third column shows the final Regotron alignments.

Although the Tacotoron2 alignment appears blurry in the middle (better viewed if zoomed in), even the very early ($13\%$) version of Regotron, has figured out how to properly align and by the time of convergence it is easy to notice that Regotron's alignment is much clearer in general. The result of Fig.~\ref{fig:alignments} is not cherry picked and should be considered representative of the alignment behavior on the held-out test set (see Regotron webpage\footnote{http://tesla.ilsp.gr:1994/regotron}). This result can be attributed to the direct penalization of the non-monotonic alignments in the attention mechanism.

\begin{table}[!htb]
  \caption{Robustness Analysis. Each kind of word error is counted at most once per sentence.}
  \label{tab:robustness_analysis}
  \centering
  \scalebox{0.9}{
  \begin{tabular}{ ccccc }
    \toprule
    \textbf{Model} & Repeat & Skip & Mispron. & Error Sent. \\
    \midrule
    Tacotron2  & $3$ & $19$ & $13$ & $54\%$ \\
    Regotron ($66\%$)  & $1$ & $10$ & $13$ & $42\%$ \\
    Regotron  & $0$ & $9$ & $10$ & $36\%$ \\
    \bottomrule
  \end{tabular}}
\end{table}

\subsection{Robustness Analysis}
The autoregressive nature of attention mechanism in both Tacotron2 and Regotron architectures, may cause
wrong attention alignments between input characters and mel-spectrogram, resulting in instability with word
repeating and word skipping. To evaluate the robustness of our model we choose two Regotron models, an earlier version ($\sim 66\%$ of the total number of training epochs) as well as the fully converged and also a Tacotron2 architecture as a baseline to compare with. 

Following FastSpeech ~\cite{ren2019fastspeech} we generated the same $50$ difficult examples 
and manually annotated them regarding three error types. \textit{Skip} errors, in which a letter/word is skipped while the nearby are uttered correctly. \textit{Mispronounciations} in which a word is pronounced incorrectly, i.e., multiple letters are skipped or confused resulting in non-interpretable speech.
\textit{Repeats} in which the same word/letter is repeated usually in place of some other. During our evaluation mispronounciations are considered a broader error class than skip errors and are evaluated separately.

Table~\ref{tab:robustness_analysis} 
results depict that the repeat errors can be fully alleviated by injecting the alignment term, which in turn shows that \
better alignment is able to tackle mistakes of this nature. The number of skip errors is also decreased but still evident, which shows that these errors are partially due to wrong alignment and partially due to model or data insufficiency. For example in phrases such as ``64x64" the intermediate ``times" or ``x" is skipped in all models examined and is uttered as if it was ``64 64". The mispronounciation errors are the most difficult to tackle as seen from
Table~\ref{tab:robustness_analysis}. Phrases s.a. ``Http", ``.dll", ``exe" or even isolated letters, e.g., "c", are not uttered correctly and the speech produced by all three models, is non-interpretable. Overall, Regotron decreases all types of errors and most importantly degrades the total error percentage, i.e., the number of sentences in which at least one error occurrs.

\begin{table}[!htb]
  \caption{MOS scores with $95\%$ confidence intervals.}
  \label{tab:MOS}
  \centering
  \scalebox{0.9}{
  \begin{tabular}{ccc}
    \toprule
    \textbf{Model} && MOS \\
    \midrule
    GT && $4.377\pm0.097$ \\
    GT (Mel + WaveGlow) && $3.983\pm0.118$ \\
    Tacotron2 (Mel + WaveGlow)  && $3.898\pm0.125$  \\
    \midrule
    Regotron$_{66\%}$ (Mel + WaveGlow) && $3.989\pm0.110$  \\
    Regotron$_{100\%}$ (Mel + WaveGlow) && $\mathbf{4.034\pm0.106}$  \\
    \bottomrule
  \end{tabular}}
\end{table}

\subsection{Speech Naturalness}
In order to assess the naturalness of the synthesized speech, we randomly selected 30 samples from our test set and used WaveGlow ~\cite{waveglow} as a vocoder. 4 different methods are evaluated, including ground truth, Tacotron2,
Regotron$_{66\%}$
and
Regotron$_{100\%}$ 
spectrograms, as well as the original speech clips. We note that 
Regotron$_{66\%}$ 
corresponds to Regotron trained 
for $66\%$ of the total training epochs
and 
Regotron$_{100\%}$
to the fully converged Regotron.
Overall we considered 150 speech samples and invited 50 evaluators. Each sample is rated by at least 5 participants on a Likert scale from 1 to 5 with 0.5 increments ~\cite{tacotron2}. 
Each evaluation test is conducted independently to avoid introducing possible bias when raters assign their subjective score.

\begin{figure}[!htb]
    \centering
    \includegraphics[width=\linewidth]{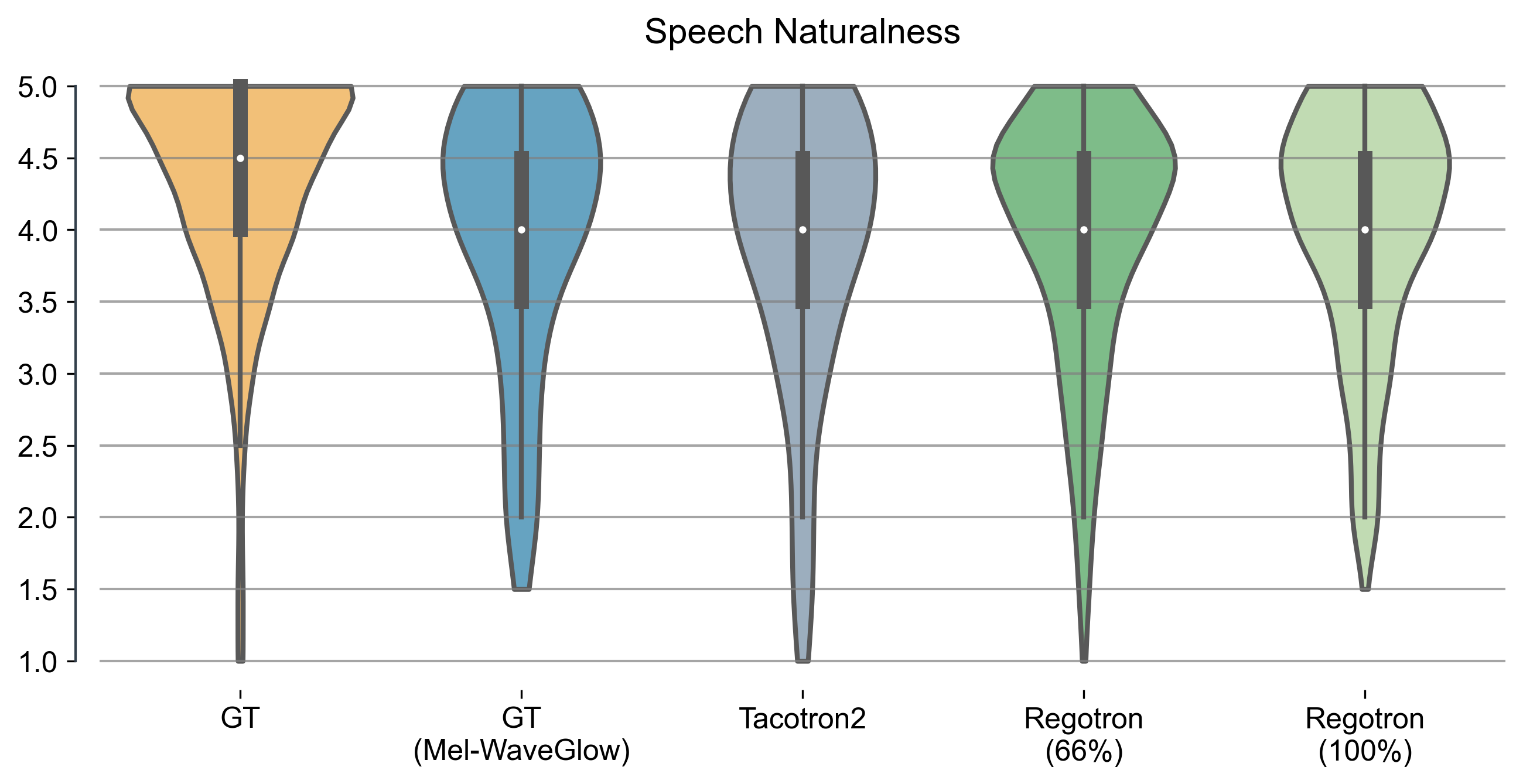}
    \caption{Boxplots of the ratings from 50 evaluators. The distributions are computed with a kernel density estimator function.}
    \label{fig:boxplots}
\end{figure}

Table~\ref{tab:MOS} presents the computed 
MOS,
indicating that Regotron$_{100\%}$ reports better MOS than Tacotron2. The evaluation also suggests that Regotron$_{100\%}$ achieves similar MOS to the speech clips that were synthesized from ground truth spectrograms. These findings are further supported by Fig.~\ref{fig:boxplots}, which shows the distributions of the collected answers.
The confidence intervals do not allow for a statistical significant conclusion, still
Regotron$_{100\%}$ distribution is similar to the one of the ground truth spectrograms (GT) and Regotron$_{66\%}$ distribution more similar to Tacotron2. This also implies that our regularization term tends to slightly improve speech naturalness.


\section{Conclusions}
This work presents Regotron, a regularized Tacotron2 variant, which aims to stabilize training and generate monotonic alignments of input text and generated mel spectrograms. We directly penalize non-monotonic alignments by adding a regularization term to the total objective function. We show that the loss curves become smoother and at the same time monotonic alignments are produced. Common TTS mistakes are reduced and a lower generalization gap is achieved. MOS is slightly improved from the addition of our term. Our framework is efficient since no additional computation is introduced.

The objective we introduce herein is generic and can be adopted by any TTS system, with potential modifications. We believe that this kind of solutions are in the pathway for tackling training stability and wrong-alignment issues in TTS systems efficiently. Applying our method to other TTS systems is very important in order to establish it as a plug'n'play method in a wide range of text-to-mel networks.



\bibliographystyle{IEEEtran}

\bibliography{arxiv}

\end{document}